\documentclass[aps,prl,twocolumn,amsmath,amssymb,superscriptaddress,citeautoscript,longbibliography,floatfix]{revtex4-1}
\newcommand{\angstrom}{\textup{\AA}}
\usepackage[allow-number-unit-breaks]{siunitx}
\usepackage{graphicx}
\usepackage{dcolumn}
\usepackage{bm}
\usepackage{color}
\usepackage{bbding}
\usepackage{pifont}
\newcommand{\xmark}{\ding{55}}
\usepackage{mathrsfs}
\usepackage{amsfonts}
\usepackage{multirow}
\usepackage{natbib}
\usepackage{epsf}
\usepackage{epsfig}
\usepackage{epstopdf}
\usepackage{amsmath}
\usepackage{hyperref}
\hypersetup{
    colorlinks=true,
    linkcolor=blue,
    filecolor=cyan,      
    citecolor=blue,
    urlcolor=blue,
    }

\begin{document}

\title{Magnetic precession induced spin accumulation in collinear antiferromagnets}

\author{Qianqian Xue}
\affiliation{Center for Alloy Innovation and Design, State Key Laboratory for Mechanical Behavior of Materials, Xi'an Jiaotong University, Xi'an, 710049, China}

\author{Jian Zhou}\email{jianzhou@xjtu.edu.cn}
\affiliation{Center for Alloy Innovation and Design, State Key Laboratory for Mechanical Behavior of Materials, Xi'an Jiaotong University, Xi'an, 710049, China}

\begin{abstract}
Generating and characterizing uniform and staggered spin polarization in antiferromagnets is one of the key challenges for antiferromagnetic spintronic technology. Here, we perform perturbative theory, group-theoretical symmetry analysis, low energy and \textit{ab initio} simulations to propose that the magnetic precession near the equilibrium magnetic axis could generate finite uniform and staggered spin polarization at the opposite magnetic sublattices (referring to total magnetic and N\'eel vector generation) in a single AFM semiconductors. This response does not require the heterojunction setup and could eliminate the lattice mismatch issues at the junction. Through scrutinizing all symmetrically-protected vanishing magnetic moment groups and especially focusing on parity-time ($\mathcal{PT}$) invariant groups, we identify the symmetry constraints that describe the staggered spin accumulation responses, and disclose their fieldlike and dampinglike characters. This unravels a hidden spin accumulation mode in AFM semiconductors. Furthermore, we simulate such an effect using a perturbative approach and suggest that electric gate field and Floquet light-dressing can effectively manipulate these responses.
\end{abstract}

\maketitle
\section{Introduction}
The modern information storage and memory largely rely on spintronics science and technology, in which the engineering and manipulation of spin polarization is of great importance. Spin pumping has emerged as one of the central mechanisms, serving as a versatile platform to explore nonequilibrium spin generation and providing a dynamic pathway to generate spin angular momentum from magnetization dynamics. Conventional spin pumping usually occurs at the heterojunction between heavy metals and ferromagnets, as the magnetic moments in ferromagnets are easier to manipulate and detect, while antiferromagnetic (AFM) systems were largely overlooked as they exhibit zero net magnetization. Recent years have witnessed a great interest on AFM spintronics owing to their immune to stray field and their ultrafast kinetics \cite{Nemec18NP,Baltz18RMP,Jungwirth16NN,Han23NM,Cheng14PhysRevLett.113.057601,Wang21PhysRevLett.127.117202,Johansen17PhysRevB.95.220408,islam2025arxiv}, yet an effective N\'eel vector polarization generation (scattered spin generation at the opposite magnetic sublattices) is one of its main obstacles for efficient information read and write technology.

In this work, we present a theoretical analysis and perform \textit{ab initio} calculations to show that magnetic precession could dynamically pump staggered spin polarization (N\'eel vector pumping) in a single semiconducting AFM system, without the need of heterojunction setup. The conventional magnetic precession dynamics generating spin polarization has been widely studied in ferromagnetic systems which are usually metallic in their conductivity. Recent efforts have been devoted into electrical toggling (staggered) spin polarizations in AFMs \cite{Han24SA,Chen25PRLzm5y-vy41,Lopez19PhysRevApplied.11.024019,Park20PhysRevB.102.224426,Ezawa25PhysRevB.111.L161301}. The contributions usually require a sizable Fermi surface with strong magnetoelectric coupling, while the Fermi sea contributions are, even though symmetrically allowed, much weaker than the Fermi surface contributions \cite{Chen25PRLzm5y-vy41}. Here, we propose another route to staggered spin polarization accumulation that could arise in intrinsic AFMs from the Fermi sea contributions. By performing perturbative response approach and magnetic group-theoretical investigations, we show that they could host a \textit{hidden} and staggered spin accumulation under magnetic precession. Our low-energy model simulation results predict that the sublattice-opposite moments in AFMs can produce a staggered spin response even when their net magnetization vanishes. Based upon magnetic group theoretical analysis, we scrutinize all $\mathcal{PT}$ invariant magnetic point groups (MPGs) and unravel the symmetric operations for the N\'eel vector. Then we elucidate the symmetry-constrained uniform and staggered spin accumulation under magnetic precession. Note that this response is different from the prior antiferromagnetic spin pumping that typically involve interfacial effects and generate spin currents \cite{PhysRevLett.113.057601,Science.368.160}. Furthermore, this is also different from the magnetoelectric coupling \cite{npj.7.2021} or spin-orbit torque \cite{Science.Advances.5.eaau6696,PhysRevLett.128.247204} that describe the electric field or current induced effective magnetic field. Here, we focus on the adiabatic torque ($\vec{\mathfrak{D}}=\hat{\bm m}\times\partial_t\hat{{\bm m}}$) induced (staggered) spin accumulation in a single phase material through perturbative responses. In order to illustrate the process, we perform first-principles calculations in a bilayer $\mathrm{MnBi}_2\mathrm{Te}_4$ and show that the sublattice-dependent spin accumulations can also be manipulated under electric and optical fields.

\begin{figure}[bt]
    \centering
    \includegraphics[width=0.4\textwidth]{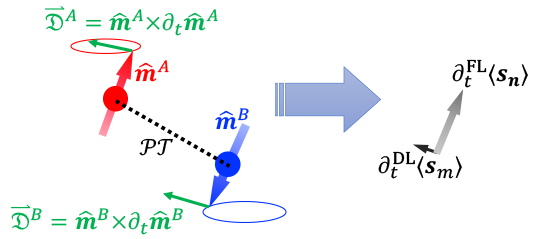}
    \caption{Schematic plot of antiferromagnetic bipartite lattice with collinear equilibrium magnetic configuration along $z$, where the magnetic precession of the two sublattices $A$ and $B$ lead to the dynamics of both total spin ($\partial_t\langle\bm s_m\rangle$) and staggered spin ($\partial_t\langle\bm s_n\rangle$) accumulation. The Gilbert damping $\vec{\mathfrak{D}}^A$ and $\vec{\mathfrak{D}}^B$ are plotted. Here, $\bm s_m=(\bm s^A+\bm s^B)/2$ and $\bm s_n=(\bm s^A-\bm s^B)/2$ are the uniform and staggered spin polarization, respectively. Here, the magnetic precession angles that deviate from the equilibrium magnetic easy axis are exaggerated for clarity reason.}
    \label{fig:Schm}
\end{figure}

\section{Computational Methods}
We perform density functional theory (DFT) calculations in the Vienna \textit{ab initio} simulation package \cite{Kresse93PRB,Kresse96PRB} within the projector augmented-wave method \cite{Blochl94PRB}. Generalized gradient approximation (GGA) method within the solid-state Perdew-Burke-Ernzerhof (PBE) form \cite{Perdew96PRL,Perdew08PRL} is adopted, and the Brillouin zone integration is performed using the Monkhorst-Pack special $k$ mesh scheme \cite{Monkhorst76PRB}. The strong correlation in the $\rm Mn$-$d$ is treated using the Hubbard $U$ method with $U=5.34\,\mathrm{eV}$. Convergence criteria of total energy and force component are set to be $1\times 10^{-7}\,\rm eV$ and $1\times 10^{-3}\,\mathrm{eV/\angstrom}$, respectively. Self-consistent spin-orbit coupling (SOC) is included within all calculations. We fit the Hamiltonian using the maximally localized Wannier functions within the Wannier90 package \cite{Mostofi14CPC}, and then the susceptibility functions are evaluated by the home-built codes. Convergences of the computational approaches have been well-tested.

\section{Results}
\subsection{Uniform and staggered spin accumulation under magnetic precession}
To describe the dynamical uniform total and staggered spin accumulation induced by collective magnetic precession, we start from the general total electronic Hamiltonian of an AFM system \cite{Haney10PhysRevLett.105.126602}
\begin{equation}
    H=H_0+J\sum_i\left[\hat{\bm m}^i(t)\cdot\bm\sigma\right],
    \label{eq:ham}
\end{equation}
where $H_0=-\frac{\hbar^2}{2m}\nabla^2+V(\bm r)$ is spin-independent part of Hamiltonian that includes kinetic and Coulomb interactions. Here, $\hat{\bm m}^i(t)$ denoting the time and site ($i$)-dependent magnetization unit vector. One could breakdown its effect into exchange and spin-orbit coupling (SOC) effect, such as $J\bm m^i(t)=(-1)^iJ_\mathrm{ex}\hat{\bm S}^i(t)/\hbar+\lambda_{\mathrm{so}}\hat{\bm L}^i/\hbar$, where $\hat{\bm S}^i(t)$ and $\hat{\bm L}^i$ being the site-dependent spin and orbital operators, respectively. $\bm\sigma$ denotes the spin-$1/2$ Pauli matrix vector. Here, we treat the exchange and SOC effects as a whole since SOC is always present in realistic materials. Then the site-dependent spin torque is \cite{Mahfouzi18PhysRevB.97.224426,Freimuth14PhysRevB.90.174423,Wimmer16PhysRevB.94.054415,Belashchenko19PhysRevMaterials.3.011401,Geranton15PhysRevB.91.014417}
\begin{equation}
    \bm T_i^{S}=\frac{1}{i\hbar}[\bm\sigma,H_i]=J\hat{\bm m}^i\times\bm\sigma.
\end{equation}
When the magnetic moment is localized near the magnetic sites, such as in $3d$ transition metal-based magnetic materials, one could show that the time-derivative of Hamiltonian is \cite{Simanek03PhysRevB.67.144418,Mills03PhysRevB.68.014419,Go25PhysRevB.111.L140409}
\begin{equation}
    \partial_tH=J\sum_i[\hat{\bm m}^i(t)\times\bm\sigma]\cdot[\hat{\bm m}^i(t)\times\partial_t\hat{\bm m}^i(t)]=\sum_i\bm T_i^S\cdot\vec{\mathfrak{D}}_i.
\end{equation}
This will be used to evaluate the spin accumulation over time. Note that here we assume a slow time variation of the magnetization precession, so that the adiabatic condition is well satisfied. The change of spin is
\begin{equation}
    \begin{split}
        \partial_t\langle S_{\alpha}\rangle=&\sum_{n,m}f_n [\langle\partial_t n|m\rangle\langle m|S_\alpha|n\rangle+\langle n|S_\alpha|m\rangle\langle m|\partial_t\ n\rangle]\\
        =&\sum_{n,m}[{f_n\langle n|S_\alpha|m\rangle\langle m|\partial_tn\rangle}+f_m\langle n|S_\alpha|m\rangle\langle\partial_tm|n\rangle]\\
        =&\sum_{n,m}{\left(f_n-f_m\right)\langle n|S_\alpha|m\rangle\langle m|\partial_tn\rangle}.
    \end{split}
\end{equation}
The last equation holds since $\partial_t\langle m|n\rangle=\langle\partial_tm|n\rangle+\langle m|\partial_tn\rangle=0$. $f_n$ refers to the Fermi-Dirac distribution of electronic structure at band $n$. We omit the $\bm k$ dependence and site index ($i$) for clarity reason. The time-derivation $|\partial_tn\rangle$ can be expressed using $\partial_tH$
\begin{equation}
    |\partial_tn\rangle=|n\rangle\left\langle n\middle|\partial_tn\right\rangle+\sum_{m\neq n}\frac{|m\rangle\langle m|\partial_tH|n\rangle}{E_n-E_m+i\eta}.
\end{equation}
Here, $E_n$ refers to the eigenenergy of band $n$ and $\eta$ is a universal phenomenological broadening factor to incorporate disorder, impurities, and environmental effects. Therefore, under magnetic precession within the adiabatic strong exchange limit, the spin accumulation is proportional to the Gilbert-damping $\vec{\mathfrak{D}}$ that serves as a generalized force for spin accumulation (Fig. \ref{fig:Schm}). The linear response perturbation formula is phenomenologically written as
\begin{equation}
    \partial_t\langle S_{\alpha}\rangle=\sum_{\alpha,\beta}\chi_{\alpha\beta}^{S}\mathfrak{\vec{D}}_{\beta}-\frac{\langle S_{\alpha}\rangle}{\tau}.
    \label{eq:generaltorque}
\end{equation}
Here, we include the phenomenological dissipation effect using an effective magnetic precession lifetime $\tau$. This indicates that the spin accumulation does not infinitely increase as time and would saturate at $\delta\langle S_{\alpha}\rangle=\tau\chi_{\alpha\beta}^S\mathfrak{D}_{\beta}$. This is similar as the injection current generation (intrinsic ballistic current) under light irradiation \cite{Sturman20PU,vonBaltz81KrautPRB,Dai23CPR,Xue23PRB}, but physically distinct from the standard SOT scenario \cite{PhysRevB.92.064415} where the torque is driven by an external electric field rather than the magnetic dynamics itself.

In this work, we mainly focus on the susceptibility function tensor $\tensor{\chi}$ perturbatively, and leave the whole magnetic dynamical effect in future works. Therefore, we explicitly emphasize that the theoretical framework in this work is established only in the adiabatic (low-frequency) regime, where the precession frequency $\omega$ is assumed to be much smaller than the characteristic electronic energy scales. More strictly, the condition $\hbar\omega \ll  E_{g}$ ($\omega$ the magnetic precession frequency) is required to suppress potential interband excitations.

The susceptibility function is a second order tensor indexed by $\alpha$ and $\beta$. Note that they are essentially denoted in the local coordinates, generated along $\partial\hat{\bm m}^i/\partial t$, $(\hat{\bm m}^i\times\partial_t\hat{\bm m}^i)$, and $\hat{\bm m}^i$. When the precession is not significantly away from the equilibrium axis (differ by a small angle $\theta\sim 0-5^{\circ}$), we approximately use the Cartesian coordinate of the equilibrium axis to conduct the analysis \cite{Go25PhysRevB.111.L140409}. In AFMs with two opposite magnetic sublattices ($A$ and $B$), the above expression in Eq. (\ref{eq:generaltorque}) can be projected onto them separately. Under a coherent magnetic precession with $\hat{\bm m}^A\simeq-\hat{\bm m}^B$, one yields $\mathfrak{\vec{D}}^A\simeq\mathfrak{\vec{D}}^B$, giving rise to site-dependent (uniform and staggered) spin polarization pumping even in the absence of a net magnetization. This condition corresponds to a phase-locked precession where the sublattice spins rotate collectively and oppositely, producing identical damping fields. This refers to the regime where our perturbation theory is formulated under the strong inter-sublattice exchange limit and within the small precession angle. In this regime, the relevant AFM eigenmode enforces an approximately antiparallel alignment, so that the relative phase between the two sublattices is locked. Deviations from such an idealized situation, such as arising from elliptical precession trajectories, unequal and potentially anisotropic damping, or large cone angles, would break the linear response treatment and are not captured by this analysis. According to the above discussion and similar as in previous works \cite{Freimuth14PhysRevB.90.174423,Go25PhysRevB.111.L140409}, the coefficient tensor $\chi_{\alpha\beta}^S$ can be evaluated via
\begin{equation}
    \begin{split}
        \chi_{\alpha\beta}^{S}=\hbar&\int[d\bm k]\left[\sum_{n,m}(f_{n\bm k}-f_{m\bm k})\right. \\
        &\left.\times\frac{\langle u_{n\bm k}|\sigma_{\alpha}|u_{m\bm k}\rangle\langle u_{m\bm k}|T_{\beta}|u_{n\bm k}\rangle}{E_{n\bm k}-E_{m\bm k}+i\eta}\right]
    \end{split}
    \label{eq:chi}
\end{equation}

If the torques on two magnetic sublattices are different, one can decompose them into uniform and staggered forms on the two magnetic sublattices, $\bm T=\bm T^A+\bm T^B$ and $\tilde{\bm T}=\bm T^A-\bm T^B$, respectively. Here, we focus on the uniform torque ($\bm T$) effect to generate non-staggered and staggered spin polarization in the main text. It dominates when $\hat{\bm m}^A\simeq -\hat{\bm m}^B$ (and $\mathfrak{\vec{D}}^A\simeq\mathfrak{\vec{D}}^B$, in-phase locked condition under strong exchange field limit). Deviations from this ideal condition, such as elliptical precession or unequal damping, can be approximately described by $\tilde{\bm T}$. Its effect on spin accumulations are briefly reported in SM \cite{supp}. In principle, it allows us to separate the precession torque effects from the individual sublattices $A$ and $B$. We emphasize that such a separation is valid only within the near-equilibrium perturbative framework, while highly nonlinear dynamical regimes lie beyond the scope of the present linear response analysis.

In Eq. (\ref{eq:chi}), the integral is performed in the $D$-dimensional first Brillouin zone, $[d\bm k]=\frac{d^D\bm k}{(2\pi)^D}$, and the value is per unit cell, volume of $V_{\mathrm{u.c.}}$. We further decompose such a response function into time-reversal $\mathcal{T}$ even and $\mathcal{T}$ odd contributions, respectively,
\begin{equation}
    \chi_{\alpha\beta}^{S,\rm even}=\hbar\int[d\bm k]\left[\sum_{n,m}(f_n-f_m)\frac{\eta\mathrm{Im}(\sigma_{nm,\alpha}T_{mn,\beta})}{(E_n-E_m)^2+\eta^2}\right]
    \label{eq:intrachi}
\end{equation}
and
\begin{equation}
    \begin{split}
        \chi_{\alpha\beta}^{S,\rm odd}=&\chi_{\alpha\beta}^{S,\rm sea}+\chi_{\alpha\beta}^{S,\rm Drude}\\
        =&\hbar\int[d\bm k]\left[\sum_{n\neq m}(f_n-f_m)\frac{\mathrm{Re}(\sigma_{nm,\alpha}T_{mn,\beta})}{E_n-E_m}\right. \\
        &\left.+\sum_n\sigma_{nn,\alpha}T_{nn,\beta}\left(\frac{\partial f_n}{\partial E}\right)\right].
    \end{split}
    \label{eq:interchi}
\end{equation}
According to the Sokhotski-Plemelj formula, $\frac{\eta}{(E_n-E_m)^2+\eta^2}\stackrel{\eta\rightarrow 0}{=}\pi\delta(E_n-E_m)\sim\tau$. Hence, the $\mathcal{T}$ odd contribution mainly arises when band $n$ and $m$ are both near Fermi level (with finite $f_n-f_m$). We also denote it as Fermi surface (sf) contribution ($\chi_{\alpha\beta}^{S,\rm sf}$) in the following discussion. This $\chi_{\alpha\beta}^{S,\rm sf}\sim\tau^1$ shows a conductive-like feature, and is potentially tuned through extrinsic environmental factors such as temperature and disorder levels. We note that the presence of dissipative broadening does not invalidate Onsager reciprocity. Instead, the response tensor can be classified according to the time-reversal parities of the generalized forces and observables. The Fermi surface and Fermi sea components may exhibit different symmetry features under magnetic-order reversal ($\chi_{\alpha\beta}(\bm m,\bm n)=\epsilon_{\alpha}\epsilon_{\beta}\chi_{\beta\alpha}(-\bm m,-\bm n)$ with $\epsilon_{\alpha}, \epsilon_{\beta} = \pm 1$ denoting the time-reversal characters of their corresponding force and observable operators), but both remain compatible with the general Onsager framework.

Furthermore, we find that in the clean limit ($\eta\rightarrow0$), the Fermi surface contributed $\chi_{\alpha\beta}^{S,\rm sf}$ is related to a ``magnetic'' Berry curvature $\chi_{\alpha\beta}^{S,\mathrm{sf}}=-2\eta\int[d\bm k]\mathrm{Im}\sum_{n}^{\mathrm{occ}}\langle\frac{\partial u_{n\bm k}}{\partial\phi_{\beta}}|\frac{\partial u_{n\bm k}}{\partial\mathfrak{J}_{\alpha}}\rangle$, where we denote the exchange field strength and magnetic unit vector as a whole vector ($\vec{\mathfrak{J}}=J\hat{\bm m}$). According to Eq. (\ref{eq:ham}), $H=H_0+\vec{\mathfrak{J}}\cdot\bm\sigma$  (the site index is omitted for clarity). One can write
\begin{equation}
\begin{split}
    \frac{\partial |u_{n\bm k}\rangle}{\partial\mathfrak{J}_{\alpha}}=&\sum_{m\neq n}\frac{|u_{m\bm k}\rangle\langle u_{m\bm k}|\frac{\partial H}{\mathfrak{J}_{\alpha}}|u_{n\bm k}\rangle}{E_{n\bm k}-E_{m\bm k}}+i\mathfrak{m}_{n\bm k}|u_{n\bm k}\rangle\\
    =&\sum_{m\neq n}\frac{|u_{m\bm k}\rangle\langle u_{m\bm k}|\sigma_{\alpha}|u_{n\bm k}\rangle}{E_{n\bm k}-E_{m\bm k}}+i\mathfrak{m}_{n\bm k}|u_{n\bm k}\rangle,
\end{split}
\end{equation}
where $\mathfrak{m}_{n\bm k}$ is a gauge-dependent real number. This measures the variation of Bloch wavefunction over exchange field components. One can artificially rotate $\hat{\bm m}$ by a small angle $\phi$. For instance, when we assume that the equilibrium magnetic axis is along $z$ and the rotation along the $x$ direction, then we have $H(\phi_x)=e^{i\sigma_x\phi_x}He^{-i\sigma_x\phi_x}$. Hence, the change of Hamiltonian with respect to this angle is
\begin{equation}
    \begin{split}
        \left.\frac{\partial H}{\partial\phi_x}\right|_{\phi_x=0}=&i\sigma_xe^{i\sigma_x\phi_x}He^{-i\sigma_x\phi_x}-ie^{i\sigma_x\phi_x}H\sigma_xe^{-i\sigma_x\phi_x}\\
        =&i[\sigma_x,H].
    \end{split}
\end{equation}
Note that $T_x=\frac{1}{i\hbar}[\sigma_x,H]=-\frac{\partial H}{\hbar\partial\phi_x}$ near the magnetic equilibrium. Similar performance can be conducted for $y$ rotation. Hence, in general one can write $T_{\beta}=-\frac{\partial H}{\hbar\partial\phi_{\beta}}$ when we introduce a vector angle $\vec{\phi}$. In this case,
\begin{equation}
\begin{split}
    \frac{\partial |u_{n\bm k}\rangle}{\hbar\partial\phi_{\beta}}=&\sum_{m\neq n}\frac{|u_{m\bm k}\rangle\langle u_{m\bm k}|\frac{\partial H}{\hbar\partial\phi_{\beta}}|u_{n\bm k}\rangle}{E_{n\bm k}-E_{m\bm k}}+i\mathfrak{t}_{n\bm k}|u_{n\bm k}\rangle\\
    =&-\sum_{m\neq n}\frac{|u_{m\bm k}\rangle\langle u_{m\bm k}|T_{\beta}|u_{n\bm k}\rangle}{E_{n\bm k}-E_{m\bm k}}+i\mathfrak{t}_{n\bm k}|u_{n\bm k}\rangle.
\end{split}
\end{equation}
Here, $\mathfrak{t}_{n\bm k}$ is another gauge-dependent real number. We note that this equation also equals to $\left(\hat{\bm m}\times\frac{\partial |u_{n\bm k}\rangle}{\partial\hat{\bm m}}\right)_{\beta}$. Therefore,
\begin{equation}
    \mathrm{Im}\langle\frac{\partial u_{n\bm k}}{\partial\mathfrak{J}_{\alpha}}|\frac{\partial u_{n\bm k}}{\partial\phi_{\beta}}\rangle=-\hbar\mathrm{Im}\frac{\langle u_{n\bm k}|\sigma_{\alpha}|u_{m\bm k}\rangle\langle u_{m\bm k}|T_{\beta}|u_{n\bm k}\rangle}{(E_{n\bm k}-E_{m\bm k})^2}.
\end{equation}
The gauge-dependent factors would vanish as they do not contribute to the imaginary parts. Therefore, the clean limit Fermi surface contribution takes the form
\begin{equation}
    \chi_{\alpha\beta}^{S,\mathrm{sf}}=-2\eta\int[d\bm k]\sum_{n}f_{n\bm k}\mathrm{Im}\langle\frac{\partial u_{n\bm k}}{\partial\mathfrak{J}_{\alpha}}|\frac{\partial u_{n\bm k}}{\partial\phi_{\beta}}\rangle.
\end{equation}
This clearly indicates that it measures a mixed ($\mathfrak{J}$-$\phi$) Berry curvature that is defined in the exchange field space, over the magnetic exchange field strength ($\mathfrak{J}$) and direction ($\phi$). The magnetic sublattice dependent responses can be brought back to describe the local mixed Berry curvature.

On the other hand, for the $\mathcal{T}$ odd susceptibility [Eq. (\ref{eq:interchi})], we find that the Drude-like intraband Fermi-surface contribution $\chi_{\alpha\beta}^{S,\rm Drude}=-\int[d\bm k]\sum_n\sigma_{nn,\alpha}T_{nn,\beta}\delta(E_n-E_F)$ is order of magnitude smaller than the rest part and can be safely omitted. This is because both $\alpha$ and $\beta$ should along the lateral directions that is normal to the equilibrium magnetization axis, which is extremely small in collinear AFMs. Hence, this $\mathcal{T}$ odd contribution is dominated by the Fermi sea contribution, which is intrinsic as $\chi_{\alpha\beta}^{S,\rm sea}\sim\tau^0$.

One can breakdown the susceptibility functions [Eqs. (\ref{eq:intrachi}) and (\ref{eq:interchi})] into different magnetic sublattice contributions, $\chi_{\alpha\beta}^{S,\mathrm{sf/sea}}=\chi_{\alpha\beta}^{S_A,\mathrm{sf/sea}}+\chi_{\alpha\beta}^{S_B,\mathrm{sf/sea}}$ for an AFM bipartite lattice. Under strong exchange condition, the uniform (staggered) local spin $\bm s_m=(\bm s^A+\bm s^B)/2$ ($\bm s_n=(\bm s^A-\bm s^B)/2$) aligns with the total magnetic moment $\hat{\bm m}$ (N\'eel vector $\hat{\bm n}$). In this regards, we define the spin difference between site-$A$ and $B$ that describes the staggered spin accumulation under magnetization precession, which are denoted as $\varsigma_{\alpha\beta}^{S,\mathrm{sf}}=\chi_{\alpha\beta}^{S_A,\mathrm{sf}}-\chi_{\alpha\beta}^{S_B,\mathrm{sf}}$ and $\varsigma_{\alpha\beta}^{S,\mathrm{sea}}=\chi_{\alpha\beta}^{S_A,\mathrm{sea}}-\chi_{\alpha\beta}^{S_B,\mathrm{sea}}$. Both variations of total and staggered spin $\delta\langle\bm s_m\rangle=\tau\chi\vec{\mathfrak{D}}$ and $\delta\langle\bm s_n\rangle=\tau\varsigma\vec{\mathfrak{D}}$ may be separately measured in AFM materials.

\subsection{Symmetry arguments}
In order to elucidate the symmetry constraints of both $\tensor{\chi}$ and $\tensor{\varsigma}$, we perform group-theoretical analysis according to Neumann's principle. We focus on collinear AFMs with symmetrically constrained zero net magnetization. For the $\mathcal{PT}$ AFMs, they conceive Kramers degeneracy in the band dispersion, as $\mathcal{PT}E_{n,\bm\sigma}(\bm k)=E_{n,-\bm\sigma}(\bm k)$ \cite{Yuan21PRM}. While the spin operator is $\mathcal{P}$ invariant ($\mathcal{P}\bm\sigma_{nm}=\bm\sigma_{nm}$) and reverses under $\mathcal{T}$ ($\mathcal{T}\bm\sigma_{nm}=-\bm\sigma_{mn}$), the N\'eel vector flips its sign under both $\mathcal{P}$ and $\mathcal{T}$. Hence, one can show that $\chi^{\mathrm{sf/sea}}$ is unchanged under $\mathcal{P}$, while $\varsigma^{\mathrm{sf/sea}}$ is forbidden. We tabulate the basic symmetry constraints in Table \ref{tab:briefPT} for $\mathcal{PT}$ AFMs. Note that the staggered spin polarization pumping depends on specific symmetry of the lattice. For example, if altermagnetic (AM, with collinear magnetization \cite{Smejkal22PRX,Smejkal22PRX-2,Zhou25PhysRevLett.134.176902}) is considered, in general the space inversion does not flip the N\'eel vector. We discuss their symmetry arguments in SM (including $\mathcal{T\bm t}$-AFMs with $\bm t$ representing half lattice translation), and focus on the $\mathcal{PT}$ AFMs in the main text.

\begin{table}[t]
    \caption{Symmetry constraints for total and staggered spin polarization susceptibility functions in $\mathcal{PT}$ AFMs. Here, the \checkmark and \xmark \,represent invariant and sign flip under transformation.}
        \centering
    \begin{tabular*}{\hsize}{@{}@{\extracolsep{\fill}}cccc@{}}
    \hline\hline
     Functions       & $\mathcal{P}$ & $\mathcal{T}$ & $\mathcal{PT}$       \\
    \hline
    $\chi^{\mathrm{sf}}$     &  \checkmark & \checkmark & \checkmark   \\
    $\chi^{\mathrm{sea}}$  & \checkmark & \xmark & \xmark \\
    $\varsigma^{\mathrm{sf}}$   &  \xmark & \checkmark & \xmark   \\
    $\varsigma^{\mathrm{sea}}$  & \xmark & \xmark & \checkmark \\
    \hline\hline
    \end{tabular*}
    \label{tab:briefPT}
\end{table}

According to Table \ref{tab:briefPT}, one observes that the $\chi^{\mathrm{sea}}$ and $\varsigma^{\mathrm{sf}}$ are always symmetrically forbidden. The former indicates that for an intrinsic AFM semiconductor without a finite Fermi surface, this magnetization dynamics induced spin polarization pumping does not appear. This is the reason that the spin pumping is usually discussed in ferromagnets rather than AFMs previously. However, when one looks at the staggered spin pumping, it only arises from Fermi sea contribution. This clearly indicates a \textit{hidden} mode for the spin polarization driven by $\mathfrak{\vec{D}}$, which requires further investigation for its emergence and tuning. The implication for deterministic N\'eel vector switching is not determined by the perturbative theory and is beyond the present linear response framework.

Next, we explore the symmetry constraints of both $\chi^{\mathrm{sf}}$ and $\varsigma^{\mathrm{sea}}$ (the other two are ignored here as they are always forbidden), we carry out magnetic group theory analysis of all the $\mathcal{PT}$ invariant systems. Note that as the torque $\bm T$ includes both the exchange and SOC contributions, in general the $SU(2)$ rotational invariant does not exist. Hence, we use magnetic group theory to conduct the analysis. In order to specifically determine how the spin and N\'eel polarization are generated under magnetic precession, we take the $2/m'$ group as an example. In this case, the equilibrium magnetization can be along the $x$, $y$, or $z$ direction. We assume that the two-fold axis is along the $z$ axis, then it contains four operators, namely, $\{E,\mathcal{C}_{2z},\mathcal{PT},\mathcal{M}_z\mathcal{T}\}$. As $\mathcal{T}$ applies on the inversion and mirror reflection, one can use $\Gamma_{\mathcal{T}}=A_u$ to represent it, according to isomorphic group method \cite{Birss64Book,Wang20npjCM,Zhou25PhysRevLett.134.176902}. The character table of $2/m$ can be seen in SM. When the equilibrium N\'eel vector is along $z$, the in-plane spin polarization transforms according to $\Gamma_{\sigma_{x,y}}=B_g$. Hence, the in-plane torque $T_{x,y}$ belongs to $\Gamma_{T_{x,y}}=B_u$ that flips its sign under $\mathcal{C}_{2z}$ and inversion $\mathcal{P}$, while keeps invariant under mirror reflection $\mathcal{M}_z$. Therefore, the $\chi_{\alpha\beta}^{S,\mathrm{sf}}$ follows $\Gamma_{\mathcal{T}}\otimes\Gamma_{\sigma_{\alpha}}\otimes\Gamma_{T_{\beta}}=A_g$ which is invariant and symmetrically allowed.

The MPGs are used to describe the magnetic order parameter $\hat{\bm m}$, which is $\mathcal{P}$ even and $\mathcal{T}$ odd. Unfortunately, it is not straightforward to describe the transformation of N\'eel vector $\hat{\bm n}$, which is both $\mathcal{P}$ and $\mathcal{T}$ odd. Hence, it is necessary to identify the groups that describe the N\'eel vector in collinear AFMs, which could then be used to determine the emergence of staggered spin pumping $\varsigma_{\alpha\beta}^{S,\mathrm{sea}}$. For instance, one can divide the operators in $2/m$ into two types, namely, $\mathcal{O}_1$ that maps the same magnetic sublattices and $\mathcal{O}_2$ between the different magnetic sublattices. For $\mathcal{PT}$ symmetric AFMs, it is evident that $E\in\{\mathcal{O}_1\}$ and $\mathcal{P}\in\{\mathcal{O}_2\}$. We denote the irreducible representation $\Gamma_{\bm n}$ that changes the sign of $\{\mathcal{O}_2\}$, yet keeping $\mathcal{O}_2\mathcal{T}$ invariant. It depends on the specific N\'eel vector $\hat{\bm n}$ direction. When $\hat{\bm n}\parallel z$, we find $\Gamma_{\bm n}=A_u$ (see SM). In this case, the N\'eel vector follows the same MPG with magnetization, $2/m'$. Following the same route, we tabulate all the MPGs that describe the N\'eel vector (denoted as $\mathcal{N}$) for all 18 $\mathcal{PT}$ invariant MPGs with collinear magnets in Table \ref{tab:MPGsymm}. It is evident that the $\mathcal{N}$'s are also black-white groups with $\mathcal{PT}$ symmetry. With all the $\mathcal{N}$ elucidated, one can further analyze the emergence of $\varsigma_{\alpha\beta}^{S,\mathrm{sea}}$. Being a $\mathcal{T}$-odd term, it simply follows $\Gamma_{n_{\alpha}}\otimes\Gamma_{T_{\beta}}$. The results are also listed in Table \ref{tab:MPGsymm}, giving a universal staggered spin accumulation in $\mathcal{PT}$ AFMs. If the system contains a high order rotation axis, then there will be only one independent component, indicating an isotropic staggered spin accumulation under adiabatic magnetic precession. We do not include $m'\bar{3}'$, $m'\bar{3}'m$, and $m'\bar{3}'m'$ in Table \ref{tab:MPGsymm} as they do not allow unidirectional magnetic pattern. With the discussions in SM \cite{supp}, we have scrutinized \textit{all} 91 MPGs that symmetrically describe both AFM and AM systems with vanishing net magnetic moments.

\begin{table}[t]
    \caption{Symmetry analysis for all $\mathcal{PT}$ invariant magnetic point groups (MPGs). $\mathrm{IR}_{\mathcal{N}}$ is the irreducible representation that swaps two magnetic sublattices. The N\'eel vector MPGs $\mathcal{N}$ are listed. Here, we assume the local coordinate as the principal rotation axis lies along $z$, and for the $m'mm$ it is the $\mathcal{M}_x$ that multiplies with $\mathcal{T}$. Note in $4'/m'm'm$ and $6'/mmm'$, the $x$ and $y$ directions of $\mathcal{N}$ are different with its parental MPG.}
        \centering
    \begin{tabular*}{\hsize}{@{}@{\extracolsep{\fill}}cccc@{}}
    \hline\hline
     MPG $\mathcal{M}$       & $\mathrm{IR}_{\mathcal{N}}$ & $\mathcal{N}$& $\varsigma_{\alpha\beta}^{\mathrm{sea}}$      \\
    \hline
    $\bar{1}'$ & $A_u$ &  $\bar{1}'$ & all allowed \\
    $2/m'$ ($\hat{\bm n}\parallel z$) & $A_u$& $2/m'$&  all allowed \\
    $2/m'$ ($\hat{\bm n}\parallel x$) & $B_u$ & $2'/m$& $\varsigma_{yz}^{\mathrm{sea}}$,$\varsigma_{zy}^{\mathrm{sea}}$ \\
    $2/m'$ ($\hat{\bm n}\parallel y$) & $B_u$ & $2'/m$& $\varsigma_{xz}^{\mathrm{sea}}$,$\varsigma_{zx}^{\mathrm{sea}}$ \\
    $2'/m$ ($\hat{\bm n}\parallel z$) & $B_u$ & $2'/m$&all allowed \\
    $2'/m$ ($\hat{\bm n}\parallel x$) & $A_u$& $2/m'$&$\varsigma_{yz}^{\mathrm{sea}}$,$\varsigma_{zy}^{\mathrm{sea}}$ \\
    $2'/m$ ($\hat{\bm n}\parallel y$) & $A_u$ & $2/m'$&$\varsigma_{xz}^{\mathrm{sea}}$,$\varsigma_{zx}^{\mathrm{sea}}$ \\
    $m'mm$ ($\hat{\bm n}\parallel x$) & $A_u$ & $m'm'm'$ &$\varsigma_{yz}^{\mathrm{sea}}$,$\varsigma_{zy}^{\mathrm{sea}}$ \\
    $m'mm$ ($\hat{\bm n}\parallel y$) & $B_{2u}$ & $mmm'$&$\varsigma_{xz}^{\mathrm{sea}}$,$\varsigma_{zx}^{\mathrm{sea}}$\\
    $m'mm$ ($\hat{\bm n}\parallel z$) & $B_{3u}$ & $mm'm$&$\varsigma_{xy}^{\mathrm{sea}}$,$\varsigma_{yx}^{\mathrm{sea}}$ \\
    $m'm'm'$ ($\hat{\bm n}\parallel x$) & $B_{3u}$ & $m'mm$&$\varsigma_{yz}^{\mathrm{sea}}$,$\varsigma_{zy}^{\mathrm{sea}}$  \\
    $m'm'm'$ ($\hat{\bm n}\parallel y$) & $B_{2u}$ & $mm'm$&$\varsigma_{xz}^{\mathrm{sea}}$,$\varsigma_{zx}^{\mathrm{sea}}$  \\
    $m'm'm'$ ($\hat{\bm n}\parallel z$) & $B_{1u}$ & $mmm'$&$\varsigma_{xy}^{\mathrm{sea}}$,$\varsigma_{yx}^{\mathrm{sea}}$ \\
    $4/m'$ ($\hat{\bm n}\parallel z$)& $A_u$& $4/m'$& $\varsigma_{xx}^{\mathrm{sea}}=\varsigma_{yy}^{\mathrm{sea}}$,$\varsigma_{xy}^{\mathrm{sea}}=-\varsigma_{yx}^{\mathrm{sea}}$  \\
    $4'/m'$ ($\hat{\bm n}\parallel z$)& $B_u$ & $4'/m'$& $\varsigma_{xx}^{\mathrm{sea}}=\varsigma_{yy}^{\mathrm{sea}}$,$\varsigma_{xy}^{\mathrm{sea}}=-\varsigma_{yx}^{\mathrm{sea}}$ \\
    $4/m'mm$ ($\hat{\bm n}\parallel z$)& $A_{1u}$ & $4/m'm'm'$&$\varsigma_{xy}^{\mathrm{sea}}=-\varsigma_{yx}^{\mathrm{sea}}$ \\
    $4'/m'm'm$ ($\hat{\bm n}\parallel z$)& $B_{2u}$ & $4'/m'm'm$&$\varsigma_{xy}^{\mathrm{sea}}=-\varsigma_{yx}^{\mathrm{sea}}$ \\
    $4/m'm'm'$ ($\hat{\bm n}\parallel z$)& $A_{2u}$ & $4/m'mm$&$\varsigma_{xy}^{\mathrm{sea}}=-\varsigma_{yx}^{\mathrm{sea}}$ \\
    $\bar{3}'$ ($\hat{\bm n}\parallel z$)& $A_u$ &$\bar{3}'$ & $\varsigma_{xx}^{\mathrm{sea}}=\varsigma_{yy}^{\mathrm{sea}}$,$\varsigma_{xy}^{\mathrm{sea}}=-\varsigma_{yx}^{\mathrm{sea}}$ \\
    $\bar{3}'m$ ($\hat{\bm n}\parallel z$)& $A_{1u}$ & $\bar{3}'m'$&$\varsigma_{xy}^{\mathrm{sea}}=-\varsigma_{yx}^{\mathrm{sea}}$ \\
    $\bar{3}'m'$ ($\hat{\bm n}\parallel z$)& $A_{2u}$ & $\bar{3}'m$&$\varsigma_{xy}^{\mathrm{sea}}=-\varsigma_{yx}^{\mathrm{sea}}$ \\
    $6'/m$ ($\hat{\bm n}\parallel z$)& $B_u$ &$6'/m$ & $\varsigma_{xx}^{\mathrm{sea}}=\varsigma_{yy}^{\mathrm{sea}}$,$\varsigma_{xy}^{\mathrm{sea}}=-\varsigma_{yx}^{\mathrm{sea}}$ \\
    $6/m'$ ($\hat{\bm n}\parallel z$)& $A_u$ & $6/m'$&$\varsigma_{xx}^{\mathrm{sea}}=\varsigma_{yy}^{\mathrm{sea}}$,$\varsigma_{xy}^{\mathrm{sea}}=-\varsigma_{yx}^{\mathrm{sea}}$ \\
    $6/m'mm$ ($\hat{\bm n}\parallel z$)& $A_{1u}$ & $6/m'm'm'$&$\varsigma_{xy}^{\mathrm{sea}}=-\varsigma_{yx}^{\mathrm{sea}}$ \\
    $6'/mmm'$ ($\hat{\bm n}\parallel z$)& $B_{2u}$ & $6'/mmm'$&$\varsigma_{xy}^{\mathrm{sea}}=-\varsigma_{yx}^{\mathrm{sea}}$ \\
    $6/m'm'm'$ ($\hat{\bm n}\parallel z$)& $A_{2u}$ & $6/m'mm$&$\varsigma_{xy}^{\mathrm{sea}}=-\varsigma_{yx}^{\mathrm{sea}}$ \\
    \hline\hline
    \end{tabular*}
    \label{tab:MPGsymm}
\end{table}

\subsection{Low energy model simulation}
\begin{figure}[bt]
    \centering
    \includegraphics[width=0.45\textwidth]{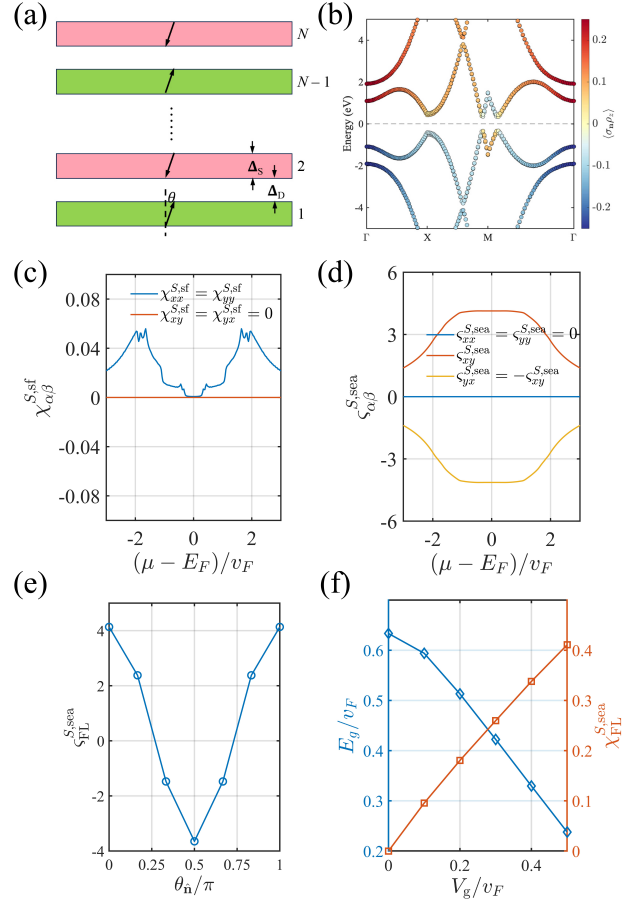}
    \caption{Low energy $\bm k\cdot\bm p$ model results. (a) Schematic plot of a layered AFM model with magnetization in each layer of polar angle of $\theta$. $\Delta_S$ and $\Delta_D$ represent interactions within each layer (top and bottom sides) and between adjacent layers. In the simulation, we assume $\theta=0$ and adopt typical parameters with $|J_{\mathrm{ex}}|=1.5v_F$, $\Delta_D=0.1v_F$, $\Delta_S=0.4v_F$, $B=0$, and $\lambda_{\mathrm{warp}}=0.1v_F$. (b) Band dispersion of a minimum bilayer AFM model with color-coding the N\'eel expectation value of each state. (c) and (d) are calculated $\chi^{S,\mathrm{sf}}$ and $\varsigma^{S,\mathrm{sea}}$ tensor components with respect to chemical potential $\mu$, respectively. (e) shows field-like $\varsigma^{S,\mathrm{sea}}$ with respect to magnetization angle $\theta$ (in the $xz$-plane) while keeping other parameters unchanged. (f) Gate voltage induced variation of band gap $E_g$ and $\chi^{S,\mathrm{sea}}$ at $\mu=-1.3v_F$.}
    \label{fig:kp}
\end{figure}

We perform numerical simulations to elucidate the magnetic precession induced dynamical local spin generation using a low energy $\bm k\cdot\bm p$ model. To set the stage, we take a layered AFM model as an exemplary platform [Fig. \ref{fig:kp}(a)]. The Hamiltonian in 2D Brillouin zone is \cite{Burkov11PhysRevLett.107.127205,Fu09PhysRevLett.103.266801,Lei20PNAS10.1073/pnas.2014004117,Li25npjCM}
\begin{equation}
    \begin{split}
        H&(\bm k)=\sum_{i,j}\bigg\{[v_F(\hat{\bm z}\times\bm \sigma)\cdot\bm k\tau_z+J_{\mathrm{ex}}^i\sigma_{\bm n}\tau_0\\
        +&(\Delta_S+Bk^2)\sigma_0\tau_x+\lambda_{\mathrm{warp}}(k_x^3-3k_x^2k_y)\sigma_0\tau_z]\delta_{i,j}\\
        +&\left(\frac{1}{2}\Delta_D\sigma_0\tau_+\delta_{j,i+1}+\frac{1}{2}\Delta_D\sigma_0\tau_-\delta_{j,i-1}\right)\bigg\}c_{\bm k,i}^\dagger c_{\bm k,j}.
    \end{split}
    \label{eq:kp}
\end{equation}
Here, $v_F$ is the Fermi velocity in the Fermi velocity for both top and bottom surfaces of each layer, indexed as $i$ ($j$). $J_{\mathrm{ex}}^i=(-1)^i|J_{\mathrm{ex}}|$ represents an alternative magnetic exchange field that points along the $\bm n$-direction. $\sigma$ and $\tau$ ($\tau_{\pm}=\tau_x\pm i\tau_y$) are Pauli matrices for spin and two surfaces of each layer, respectively ($\sigma_{\bm n}=\bm\sigma\cdot\hat{\bm n}$). $\Delta_D$ and $\Delta_S$ measure the interlayer and intralayer interactions, respectively, and only the nearest neighbor surface interactions are accounted. $\lambda_{\mathrm{warp}}$ is the hexagonal warping SOC strength. For a bilayer system, the calculated band dispersion is plotted in Fig. \ref{fig:kp}(b) with a band gap of $E_g=0.67v_F$, in which different color represents expectation value of N\'eel vector $\langle\sigma_z\rho_z\rangle$ with $\rho_z$ representing a Pauli-$z$ matrix for layer pseudospin index. With finite SOC, the N\'eel vector value is continuously changed, being not a good quantum number.

Our simulated interband and intraband uniform and staggered spin accumulations are in fully agreement with previous symmetry constraints. In detail, one sees that both $\chi^{S,\mathrm{sea}}$ and $\varsigma^{S,\mathrm{sf}}$ totally vanish, and the $\chi^{S,\mathrm{sf}}$ and $\varsigma^{S,\mathrm{sea}}$ with respect to chemical potential $\mu$ are depicted in Figs. \ref{fig:kp}(c) and \ref{fig:kp}(d). The diagonal terms of $\chi^{S,\mathrm{sf}}$ becomes nonzero when the chemical potential crossing either valence or conduction bands, while the off-diagonal terms remain to be vanishing throughout the energy window. As for the $\varsigma^{S,\mathrm{sea}}$, its diagonal components are completely zero, and the off-diagonal terms exhibit antisymmetric behaviors ($\varsigma^{S,\mathrm{sea}}_{xy}=-\varsigma^{S,\mathrm{sea}}_{yx}$). These facts are consistent with the symmetry constraints. In the following, we refer to the diagonal and off-diagonal terms the dampinglike (DL, $\propto\hat{\bm m}\times(\hat{\bm m}\times\bm\sigma)$) and fieldlike (FL, $\propto\hat{\bm m}\times\bm\sigma$) responses, respectively \cite{Go25PhysRevB.111.L140409}.

We find that the equilibrium N\'eel direction $\hat{\bm n}$ would affect the responses. In Fig. \ref{fig:kp}(e), we rotate $\hat{\bm n}$ in the $xz$-plane (keeping the azimuthal angle $\varphi_{\bm n}=0$) and plot the $\varsigma^{S,\mathrm{sea}}_{\mathrm{FL}}$ variation. Other $\varphi_{\bm n}$ values do not significantly affect the results. We observe that the $\varsigma^{S,\mathrm{sea}}_{\mathrm{FL}}$ reduces from positive values to negative values when the equilibrium $\theta_{\bm n}$ rotates from $0$ (out-of-plane) to $\pi/2$ (in-plane), and then increases back at $\pi$.

We further propose that one can apply a gate voltage to break $\mathcal{PT}$ and trigger the emergence of $\chi^{S,\mathrm{sea}}$ and $\varsigma^{S,\mathrm{sf}}$. Here, we introduce $H_V=V_g\sigma_0\tau_0\rho_z$ that generates an potential gradient at different magnetic sublattices. Figure \ref{fig:kp}(f) shows the variation of band gap $E_g$ and $\chi_{\mathrm{FL}}^{S,\mathrm{sea}}$. Under such an electric Stark effect, the former gradually reduces, while the latter increases monotonously. These provide efficient approaches to modulate the spin and N\'eel dynamics under magnetic precession. We would like to remark that the above discussions are not constrained in layered-AFMs. One can refer to results of other type AFMs in SM \cite{supp}. This is different from usual electric controlling magnetic responses in multiferroic systems \cite{Hur04Nature,Xiang08PhysRevLett.101.037209,Iniguez08PhysRevLett.101.117201,Bousquet11PhysRevLett.107.197603,Ye14PhysRevB.89.064301,Zhai17NCs41467-017-00637-x,Yamauchi19PRB,Jiang18NM}, as here we only adopt conventional AFMs. Thus, breaking $\mathcal{PT}$ symmetry restores previously forbidden uniform and staggered channels, enabling tunable dynamical responses.

\subsection{\textit{Ab initio} calculations on realistic even layer $\rm MnBi_2Te_4$}
\begin{figure}[bt]
    \centering
    \includegraphics[width=0.45\textwidth]{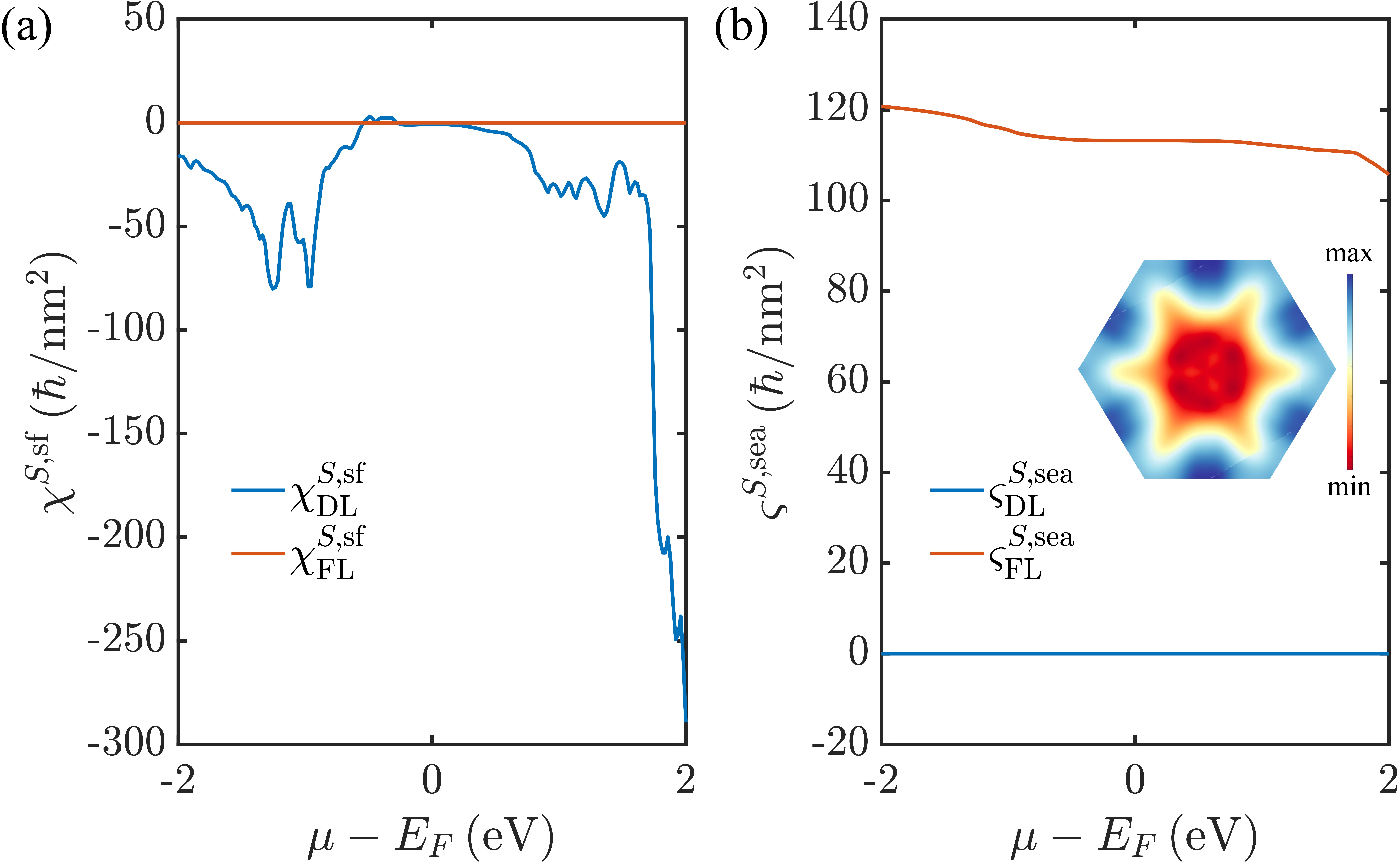}
    \caption{Results for a $\mathrm{Mn}\mathrm{Bi}_2\mathrm{Te}_4$ bilayer with equilibrium out-of-plane magnetization. (a) $\chi^{S,\mathrm{sf}}$ and (b) $\varsigma^{S,\mathrm{sea}}$ for different chemical potential. The BZ-resolved $\varsigma_{\mathrm{FL}}^{S,\mathrm{sea}}$ at the Fermi level is shown as inset plot.
    }
    \label{fig:mbt}
\end{figure}

Armed with low energy model results, we perform \textit{ab initio} calculations for a realistic material example, $\mathrm{MnBi}_2\mathrm{Te}_4$ with even septuple layers. We consider the magnetic axis along the $z$ direction, giving a MPG of $\bar{3}'m'$. With the strong correlation of $\mathrm{Mn}$-$d$ orbitals treated by the Hubbard $U$ corrections, the calculated bandgap of bilayer $\mathrm{MnBi}_2\mathrm{Te}_4$ is $E_g=0.076\,\mathrm{eV}$. Figure \ref{fig:mbt} shows our simulation results. The dampinglike total spin generation $\chi_{\mathrm{DL}}^{S,\mathrm{sf}}$ reaches $-63.9\,\hbar/\mathrm{nm}^{2}$ when $\mu=-1\,\mathrm{eV}$ below the intrinsic Fermi level, while the fieldlike $\chi_{\mathrm{FL}}^{S,\mathrm{sf}}$ is completely silent according to symmetry constraints [Fig. \ref{fig:mbt}(a)]. The susceptibility function is measured in unit of $\hbar/\mathrm{nm}^{2}$, indicating a steady-state angular momentum accumulation rate per 2D area. From Table \ref{tab:MPGsymm}, its N\'eel vector follows $\mathcal{N}=\bar{3}'m$, giving only one independent fieldlike $\varsigma_{\mathrm{FL}}^{S,\mathrm{sea}}(=\varsigma_{xy}^{S,\mathrm{sea}}=-\varsigma_{yx}^{S,\mathrm{sea}})$. This Fermi sea contribution is notably large compared to the Fermi sea part in conventional spin-orbit torque setups, suggesting it is potentially within reach of experimental detection conditions. As shown in Fig. \ref{fig:mbt}(b), in the intrinsic bandgap one has $\varsigma_{\mathrm{FL}}^{S,\mathrm{sea}}$ reaches $113.2\,\hbar/\mathrm{nm}^{2}$. Its BZ-resolved contribution is plotted in the inset, giving a clear hexagonal pattern and the main contributions from the states with large interlayer mixing near $M$ point. We provide an order of magnitude estimate of the N\'eel vector accumulation. Taking a typical ferromagnetic resonance precession frequency of $\omega_0/2\pi=10\,\rm GHz$, and a small cone angle of $\theta\sim 1^{\circ}$, the adiabatic damping vector magnitude is then $|\mathfrak{D}|=\omega_0\sin\theta\simeq 1\times 10^9\,\rm rad/s$. This yields a steady-state sublattice $\delta\langle \bm s_{n}\rangle$ on the order of $10^{-3} \,\hbar/\rm nm^2$ (taking a typical relaxation lifetime of $10\,\rm fs$). We emphasize that this is only an order estimate, but it could be within the sensitivity range of current magneto-optical or spin-resolved probes.

\begin{figure*}[bt]
    \centering
    \includegraphics[width=0.85\textwidth]{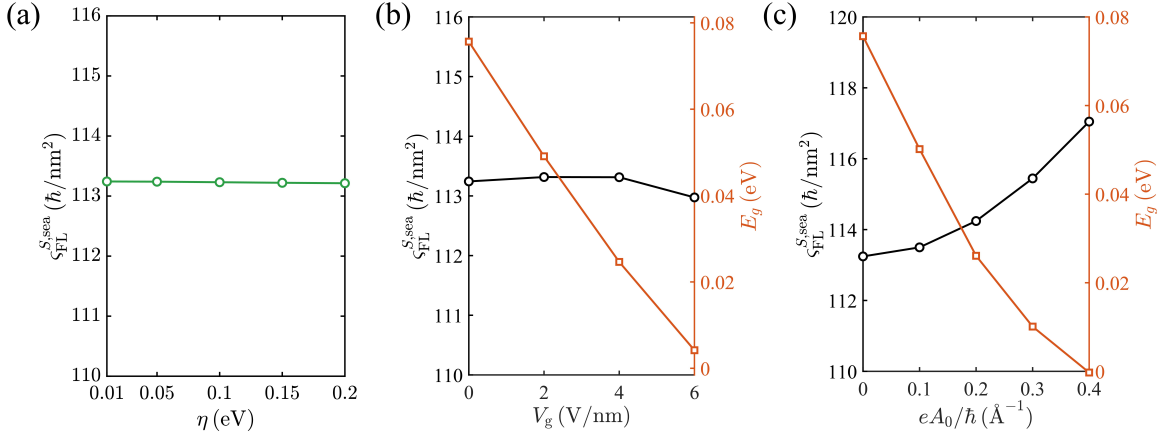}
    \caption{Field effect to modulate the $\varsigma_{\mathrm{FL}}^{S,\mathrm{sea}}$ in bilayer $\mathrm{MnBi_2Te_4}$. (a) The dependence of the in-gap $\varsigma_{\mathrm{FL}}^{S,\mathrm{sea}}$ on the phenomenological scattering rate $\eta$. (b) Gate electric field induced variation of bandgap $E_g$ (red squares, right axis) and in-gap $\varsigma_{\mathrm{FL}}^{S,\mathrm{sea}}$ (black circles, left axis). (c) Floquet light-dressed variation of $E_g$ and in-gap $\varsigma_{\mathrm{FL}}^{S,\mathrm{sea}}$ as a function of light field intensity $eA_0/\hbar$ under a left-handed circularly polarized light with pump frequency $\omega=5\,\mathrm{eV}$.}
    \label{fig:tuning}
\end{figure*}

To further illustrate the universality and robustness of this N\'{e}el accumulation in such an intrinsic semiconductors, we focus on the in-gap responses under external fields and disorder scattering on the bilayer $\mathrm{MnBi_{2}Te_{4}}$. We first examine the dependence of the response on the scattering rate $\eta$, which semiempirically incorporates the disorder level and temperature effect. Figure \ref{fig:tuning}(a) shows that the in-gap $\varsigma_{\mathrm{FL}}^{S,\mathrm{sea}}$ remains constant over a wide range of $\eta$ (from $0.01$ to $0.2\,\mathrm{eV}$). It is noteworthy that this robustness persists even when the scattering energy scale exceeds the intrinsic band gap ($E_g=0.076\,\mathrm{eV}$). This persistent stability against phenomenological broadening further corroborates the intrinsic deep sea nature of the generated staggered spin polarization.

Secondly, similar as in the model calculations, we consider a gate electric field effect $V_{g}$ along the out-of-plane direction, added through the saw-tooth electric potential method. As illustrated in Fig. \ref{fig:tuning}(b), the bandgap $E_{g}$ decreases drastically as $V_{g}$ gradually applied. Interestingly, the fieldlike staggered spin generation $\varsigma_{\mathrm{FL}}^{S,\mathrm{sea}}$ within the bandgap also does not change too much as the voltage increases. It maintains a sizable value with only marginal fluctuations even when the gap nearly closes. This distinction implies that while the electric field effectively modulates the band-edge spectrum, the induced N\'eel polarization $\varsigma$ arises primarily from the global topology of the deep Fermi sea. Consequently, it is topologically protected and insensitive to the local bandgap modulation, unlike the surface-sensitive or symmetry-breaking dependent responses.

Furthermore, we propose that a light field with ultrafast periodic time dressing could serve as an effective knob to manipulate the band dispersion and staggered spin generations. Here, we simulate such an effect via the Floquet band engineering \cite{Eckardt17RevModPhys.89.011004,Shirley65PhysRev.138.B979,Sambe73PhysRevA.7.2203,Dunlap86PhysRevB.34.3625,Zhan24QFs44214-024-00067-z,Zhou24NL} under a left-handed circularly polarized light field, $A (t)=A_{0}(\cos \omega t, \sin \omega t, 0)$. The pump frequency $\omega = 5$ eV is chosen to be significantly higher than the characteristic time scales of the relevant band energy and magnetic precession, allowing the application of the van Vleck-Magnus expansion to approximate the light dressed Hamiltonian, $H^{F} = H_{0}+\sum_{n\ge1} \frac{[H_{-l},H_{l}]}{l\hbar\omega} +\mathcal{O}(\frac{1}{\omega^{2}})$. Here, $H_{l} = \frac{1}{2\pi} \int_{0}^{2\pi/\omega} H(k+ \frac{eA(t)}{\hbar} )e^{il\omega t}dt$ denotes the $n$-th Fourier component of the time-periodic Hamiltonian, and we retain terms up to $l = 2$ to ensure convergence. As shown in Fig. \ref{fig:tuning}(c), as the light intensity enhances, the bandgap of bilayer $\mathrm{MnBi_{2}Te_{4}}$ gradually reduces and closes at $eA_{0} / \hbar \approx 0.4\,\angstrom^{-1}$. Further increasing light intensity would drive a topological phase transition \cite{Zhou24NL}, yet is challenging to be achieved experimentally. Concomitantly, the intrinsic staggered spin pumping $\varsigma_{\mathrm{FL}}^{S,\mathrm{sea}}$ within the bandgap shows a monotonic increase. This indicates that light field could serve as an effective scheme to tune the staggered spin polarization pumping in intrinsic semiconducting AFMs.

\section{Discussion and conclusion}
In addition to spin pumping, recent years have witnessed a lot of researches on the orbital accumulation and generation, dubbed orbital pumping (and orbital transport) in orbitronic field \cite{Bernevig05PRL,Jo24npjSpin,Cysne21PRL,Go23PhysRevLett.130.246701,Han25PhysRevLett.134.036305,Go20PRResearch,Mu25PhysRevB.111.165102}. This site-dependent hidden spin pumping can be extended to orbital contributions, especially when localized orbitals are within consideration (such as van der Waals layered structures). Because the torque operator couples equally to spin and orbital angular momentum, the same formalism yields a staggered orbital pumping tensor upon replacing local $\hat{\bm\sigma}$ by $\hat{\bm L}$.

In this work we are focusing on the perturbative response theory and symmetry constraint analysis. Here, the magnetic torque is considered as a source to generate staggered spin polarization, which does not correspond to the Landau-Lifshitz-Gilbert equation based magnetic dynamics for N\'eel control. Experimentally, the dynamically generated staggered spin accumulation may be detected by measuring spin changes within the AFM sample, or by attaching a heavy metal layer, where the resulting spin signal could be converted into a measurable charge current or voltage through spin-charge conversion processes such as inversion spin Hall and inverse spin Edelstein effect. While the present theoretical framework and calculations establish the symmetry conditions and microscopic mechanisms for the proposed pumping response, a quantitative estimate of experimentally detectable signal magnitudes would additionally require realistic modeling of dynamical precession amplitudes, damping, and device-specific conversion efficiencies. Such quantitative analysis is beyond the scope of the present perturbative treatment and will be investigated in future work.

In summary, we propose the hidden staggered spin accumulation under magnetic precession in AFM systems. We show how the Fermi surface and Fermi sea contributions are symmetrically constrained. We develop a magnetic group-theoretical method to elucidate the MPG that $\hat{\bm n}$ follows. The Fermi surface contributions arises from multiple bands, and scales with a mixed Berry curvature in the magnetization space. Furthermore, our theory are illustrated using low energy $\bm k\cdot\bm p$ model and \textit{ab initio} simulations for $\mathcal{PT}$-symmetric AFMs. The generation of N\'eel polarizations can be effectively controlled by light field, yet is robust against extrinsic scattering factors.

\begin{acknowledgments}
\textit{Acknowledgments.} J.Z. acknowledges useful discussions with C. Niu. This work is supported by the National Natural Science Foundation of China (NSFC) under Grant No. 12374065.
\end{acknowledgments}

\textit{Data availability.} The data that support the findings of this article are not publicly available yet. The data are available from the authors upon reasonable request.


\providecommand{\noopsort}[1]{}\providecommand{\singleletter}[1]{#1}%

\end{document}